\documentclass[prd,nofootinbib,preprintnumbers,floatfix]{revtex4}  % for review and submission
\usepackage[plainpages=false, colorlinks=true, anchorcolor=blue, linkcolor=blue, citecolor=blue, bookmarks=false]{hyperref}

\usepackage{amsmath}
\usepackage{graphicx}
\usepackage{graphicx}% Include figure files
\usepackage{dcolumn}% Align table columns on decimal point
\usepackage{bm}
\usepackage{xcolor}
\usepackage{tabularx}

\usepackage{newtxtext,newtxmath}
\usepackage[T1]{fontenc}

\newcommand{\vc}[1]{\textbf{\em #1}}

\newcommand{\pder}[2]{\frac{\partial #1}{\partial #2}}
\usepackage{adjustbox}
\begin{document}

%\preprint{APS/123-QED}

\title{Particle acceleration in self-driven turbulent reconnection}


\author{Jian-Fu Zhang$^{1}$}\email{E-mail: jfzhang@xtu.deu.cn}
\author{Siyao Xu$^{2}$} \email{E-mail: sxu@ias.edu}
\author{Alex Lazarian$^{3,4}$}
\author{Grzegorz Kowal$^{5}$}%

\affiliation{$^{1}$ Department of Physics, Xiangtan University, Xiangtan 411105, China}
\affiliation{$^{2}$ Institute for Advanced Study, 1 Einstein Drive, Princeton, NJ 08540, USA}
\affiliation{$^{3}$ Department of Astronomy, University of Wisconsin, 475 North Charter Street, Madison, WI 53706, USA}
\affiliation{$^{4}$ Centro de Investigación en Astronomía, Universidad Bernardo O’Higgins, Santiago, General Gana 1760, 8370993, Chile}
\affiliation{$^{5}$ Escola de Artes, Ci\^encias e Humanidades, Universidade de S\~ao Paulo, Av. Arlindo B\'ettio, 1000 -- Vila Guaraciaba, CEP: 03828-000, São Paulo - SP, Brazil}

\date{\today}% It is always \today, today,

\begin{abstract}
The theoretical prediction that magnetic reconnection spontaneously drives turbulence has been supported by magnetohydrodynamic (MHD) and kinetic simulations. While reconnection with externally driven turbulence is accepted as an effective mechanism for particle acceleration, the acceleration during the reconnection with self-driven turbulence is studied for the first time in this work. By using high-resolution 3D MHD simulations of reconnection with self-generated turbulence, we inject test particles into the reconnection layer to study their acceleration process. We find that the energy gain of the particles takes place when they bounce back and forth between converging turbulent magnetic fields. The particles can be efficiently accelerated in self-driven turbulent reconnection with the energy increase by about 3 orders of magnitude in the range of the box size. The acceleration proceeds when the particle gyroradii become larger than the thickness of the reconnection layer. We find that the acceleration in the direction perpendicular to the local magnetic field dominates over that in the parallel direction. The energy spectrum of accelerated particles is time-dependent with a slope that evolves toward -2.5. Our findings can have important implications for particle acceleration in high-energy astrophysical environments.

\end{abstract}

\maketitle

\section{Introduction}\label{intro}
Almost all observed high-energy astrophysical processes indicate the existence of high-energy cosmic ray particles \citep[e.g.,][]{Longair:2011,Lhaaso:2021}, the origin of which is one of the most important but unsolved problems in space physics and astrophysics. From a theoretical perspective, theories of three classical acceleration mechanisms, including stochastic acceleration (second-order Fermi), diffusive shock acceleration (first-order Fermi), and magnetic reconnection acceleration, have been developed to understand the observational phenomena in the solar wind and various astrophysical environments \citep[see reviews by, e.g.,][]{Lazarian_etal:2020,Marcowith:2020,Liu:2021FrASS}. The essence of a particle's acceleration is the interaction between (turbulent) magnetic fields and particles, such as gyroresonant interaction and transit time damping \citep{Kulsrud:1969,Schlickeiser:2002,Yan:2004,Xu:2018b,Zhdankin:2022}, mirror interaction \citep{Lazarian:2021} and other non-resonant interactions \citep{BrunettiLazarian:2016,Xu:2017,Comisso:2022,Bresci:2022,Huang2012}, gradient drift and curvature drift in non-uniform magnetic fields \citep{Caprioli:2014,Marcowith:2020,Guo:2021}. This work focuses on the magnetic reconnection acceleration of particles in the presence of turbulence that has been actively studied in recent years \citep[e.g.,][hereafter LV99]{LazarianVishniac:1999} and \citep{deGouveiaLazarian:2005,Kowal_etal:2009,Beresnyak:2016,Li:2019,Chitta:2020,Jafari:2021,Kadowaki:2021,ZhangQL:2021,ZhangSironi:2021,Xu:2023}.

The theoretical studies of magnetic reconnection date back to the 1950s \citep{Sweet:1958,Parker:1957,Petschek:1964}, with intensive efforts devoted to explaining the reconnection rates indicated by observations \citep[see][for a recent review]{Lazarian_etal:2020}. Appealing to the ubiquitous turbulence as a trigger and regulator
 of the reconnection process, LV99 proposed a 3D turbulent reconnection model describing a fast reconnection process with a reconnection rate independent of micro-plasma effects. The fast reconnection of turbulent magnetic fields can lead to a more efficient transfer of magnetic energy to the kinetic energy of the fluid. These predictions have been confirmed by both non-relativistic \citep{Kowal_etal:2009,Kowal_etal:2017,Kowal_etal:2020} and relativistic \citep{Takamoto_etal:2015}
MHD turbulence simulations. The LV99 model 
makes the modern MHD turbulence theory \citep[][hereafter GS95]{GoldreichSridhar:1995}
self-consistent and predicts the rate of magnetic field diffusion in space that is consistent with simulations \citep{Lazarian:2004,Beresnyak:2013}. 
Very importantly, turbulent reconnection predicts that the classical magnetic flux-freezing theorem \citep{Alfven:1942} is violated in turbulent fluids \citep{Eyink:2011}. 

Various astrophysical implications of the LV99 reconnection are reviewed in \citep{Lazarian_etal:2020}, including anomalous cosmic rays \citep{LazarianOpher:2009}, nonlinear turbulent dynamo \citep{XuLazarian:2016}, (first) star formation 
\citep{Stacy:2022}, gamma-ray bursts \citep{Lazaria_etal:2003,ZhangYan:2011,LazarianZhangXu:2019}, microquasars \citep[][henceforth GL05]{deGouveiaLazarian:2005}, active galactic nuclei \citep[AGNs:][]{Kadowaki_etal:2015,KhialideGouveia:2016} and radio galaxies \citep{BrunettiLazarian:2016}. By studying emissions from accelerated particles, one can get insight into the processes of turbulent reconnection.

Earlier studies on reconnection acceleration mostly focus on 2D reconnection models. \cite{Drake_etal:2006} considered the 2D reconnection and reported that the first-order Fermi acceleration can happen in this case. In the restricted 2D configuration, 
particles are trapped within contracting magnetic islands, i.e., plasmoids, which can arise from the plasma tearing instability when a long narrow current sheet is prescribed \citep[see][]{Loureiro_etal:2007}. 
The trapping of particles within the islands limits the 
acceleration efficiency in 2D reconnection. 
Moreover, 2D magnetic islands are not tenable and susceptible to turbulence (given a sufficiently large system size) in 3D, where 3D open-ended loops are expected. 
As shown by MHD and kinetic simulations  \citep[e.g.,][]{Cargill:2012,Li:2019,Vlahos:2019,ZhangQL:2021,ZhangSironi:2021}, 3D reconnection acceleration is more efficient than their 2D counterpart. 
In analogy to the classical diffusive shock acceleration (first-order Fermi), where particles are confined in the vicinity of a shock,
GL05 first proposed that a first-order Fermi process operates within the 3D turbulent reconnection region. Particles are confined in the converging magnetic fluxes of opposite polarity \citep[e.g., see Fig. 1 in][]{Nowak:2021} and bounce back and forth between the reconnection-driven inflows.
A more rigorous study on turbulent reconnection acceleration is recently carried out by \citep{Xu:2023}. They found an efficient acceleration in the presence of reconnection-driven turbulence and non-universal particle energy spectral indices depending on the reconnection rate. We note that, unlike the shock acceleration, fluid compressibility \citep{Drury:2012} is not necessary for reconnection acceleration to happen. 
The testing of the GL05 picture was performed by \citep{Kowal_etal:2011} using 3D turbulent reconnection simulations
\citep{Kowal_etal:2009}. \citep{Kowal_etal:2011} provided the first numerical demonstration of the volume-filling reconnection that happens in the 3D case. They further showed that the reconnection acceleration is efficient in the presence of turbulence.

The injection mechanism of turbulence varies in different astrophysical environments, but in general, it can be divided into externally driven turbulence and spontaneously driven one. For instance, supernova explosions \citep{NormanFerrara:1996,Ferriere:2001}, merger events and AGN feedback \citep{Chandran:2005,Subramanian_etal:2006,EnsslinVogt:2006} and baroclinic forcing behind shock waves \citep{Inoue:2013,HuXU:2022,Dhawalikar:2022} are frequently also quoted by the sources of turbulence. In addition, various instabilities such as magnetorotational instability in accretion disks \citep{BalbusHawley:1998} and kink instability of twisted flux tubes in the solar corona \citep{GerrardHood:2003} can also excite turbulence. Particularly, turbulence can be driven by magnetic reconnection itself as suggested in LV99 and \citep{LazarianVishniac:2009}. Numerical studies on reconnection-driven turbulence are very challenging because they require high resolution
and a long computational time to reach a steady state. Attempts including \citep{Beresnyak:2017} for incompressible medium, and  \citep{Oishi_etal:2015} and \citep{HuangBhattacharjee:2016} for compressible medium demonstrate the generation of turbulence by reconnection.

The dynamical evolution and statistical properties of reconnection-driven turbulence (initiated by stochastic noise) were studied by \cite{Kowal_etal:2017}. They found that the reconnection produces a Kolmogorov-like spectrum of velocity fluctuations with the anisotropic scaling following the relation $l_{\parallel} \propto l_{\perp}^{2/3}$ predicted in GS95, where $l_{\parallel}$ and $ l_{\perp}$ are the parallel and perpendicular scales of a turbulent eddy with respect to the local magnetic field.\footnote{The notion ``local" means that the scaling of turbulent eddies is measured with respect to the local magnetic field in the direct vicinity of these eddies, rather than with respect to the global mean magnetic field. This notion was absent in the original GS95 treatment, but introduced later in LV99 and \cite{Cho:2000}. The actual notion of eddies in strong Alfv\'enic turbulence is based on the LV99 finding that the turbulent reconnection takes place over just one turnover of eddies.} 

Furthermore, \cite{Kowal_etal:2020} found that
the Kelvin-Helmholtz instability dominates over the tearing instability for the generation of turbulence in the 3D reconnection layer, while the tearing instability is only important at the initial stage of the reconnection. In the absence of external driving, 
the reconnection-driven turbulence in 3D enables fast reconnection.

Our goal is to numerically study particle acceleration in 3D magnetic reconnection with reconnection-driven turbulence, which has not been numerically investigated. By performing high-resolution MHD simulations,
we will examine the effect of reconnection-driven turbulence on reconnection acceleration and acceleration efficiency.
In Section \ref{SimMeth}, we describe the numerical methods for reconnection simulations with reconnection-driven turbulence and test particle simulations. Section \ref{Results1} presents our numerical results for reconnection acceleration. Finally, we provide a discussion in Section \ref{Diss} and a summary in Section \ref{Summ}.

\section{Simulation Methods}\label{SimMeth}
\subsection{Reconnection simulations with reconnection-driven turbulence}\label{Recdriturb}
A high-order shock-capturing Godunov-type code AMUN\footnote{https://bitbucket.org/amunteam/amun-code/} is adopted to solve a set of equations for isothermal compressible magnetohydrodynamics
\begin{eqnarray}
 \pder{\rho}{t} + \nabla \cdot \left( \rho \vc{v} \right) = 0,  \ \ \ 
 \label{eq:mass} \\
 \begin{aligned}
 \pder{\rho \vc{v}}{t} + \nabla \cdot \left[ \rho \vc{v} \vc{v} +
 \left( c_{\rm s}^2 \rho + \frac{B^2}{8 \pi} \right) I - \frac{1}{4 \pi} \vc{B} \vc{B}
 \right] = & \\ \nu \rho \left[ \nabla^2 \vc{v} + \frac{1}{3} \nabla \left( \nabla \cdot \vc{v} \right) \right], & \end{aligned} \  \  \label{eq:momentum}  \\
\pder{\vc{B}}{t} + \nabla \times \vc{E} = 0, \ \ \  \label{eq:induction}\\  \nabla \cdot \vc{B} =0, \ \  \  \label{eq:non-div}
\end{eqnarray}
where $\rho$ is the plasma density, $\vc{v}$ the fluid velocity, $\vc{B}$ the magnetic field, $c_{\rm s}$ the isothermal sound speed, $\nu$ the viscosity coefficient, and $I$ the identity matrix. With Ohm's law, the electric field can be written as 
\begin{equation}
\vc{E} = - \vc{v} \times \vc{B} + \eta \, \vc{J}, \label{eq:elec}
\end{equation}
 where $\vc{J} = \nabla \times \vc{B}$ is the current density and $\eta$ is the Ohmic resistivity coefficient. In addition, an isothermal equation of state is used to supplement the above equations. The MHD equations (\ref{eq:mass})-(\ref{eq:non-div}) were integrated numerically using 3$^{th}$-order 4-stage Strong Stability Preserving Runge-Kutta method \citep{Gottlieb_etal:2011}, 7$^{th}$-order Compact Monotonicity Preserving interpolation \citep{SureshHuynh:1997} to reconstruct the Riemann states, and the HLLD Riemann solver \citep{Mignone:2007}.
 
Our numerical simulations are performed in a 3D domain with physical dimensions of $L_x=L$, $L_y=4L$, and $L_z=L$ ($L=1$), fixing the domain center at the origin of Cartesian coordinates. We consider periodic boundary conditions along the $X$ and $Z$ axes and an open one along the $Y$ axis. Uniform initial density is set to be $\rho=1$ in the upstream region, and within the current sheet, its value increases to make the total pressure uniform in the entire domain. The initial antiparallel magnetic field with the strength of $B_0=1$ is set along the $X$ axis, with a single discontinuity of the $X$-component of the magnetic field placed in the middle of the box on the $X$-$Z$ plane, and the guide field strength of $B_{\rm g}=0.1$ along the $Z$ axis.\footnote{This work is limited to a single guide field strength. Note that recent kinetic scale simulations have claimed that the guide field plays a critical role in the reconnection acceleration of particles  \citep{Dahlin:2015,Li:2017,Arnold:2021}. However, MHD-scale turbulent reconnection simulations claimed that the contribution of the guide field does not change the dynamics of magnetic reconnection and its reconnection rate \citep{Beresnyak:2012,Beresnyak:2017}. More confirmation is needed.} These settings result in the dimensionless fluid velocity and the simulation time units to be $[ v ] = V_{\rm A} = 1$ and $[t]=t_{\rm A} = L / V_{\rm A} =1$, respectively. Since the AMUN code runs with the normalized magnetic field multiplied by a factor $\sqrt{4 \pi}$, Alfv\'en velocity using the code units is $V_A = B_0 / \sqrt{\rho_0}$. Furthermore, the initial random velocity field with an amplitude of $10^{-2}$ at the fixed wavenumber of $k = 32 \pi$ is represented by $100$ Fourier modes of random phases and directions, being limited in the spatial region within the distance of $0.1$ from the initial magnetic field discontinuity \cite[see][for more details]{Kowal_etal:2017}. Besides, the isothermal sound speed is set to be $c_{\rm s} = 1$, resulting in a plasma parameter of $\beta = p_{\rm g} / p_{\rm m} \approx 2$. To control the effects of numerical diffusion, we set the viscosity $\nu$ and resistivity $\eta$ coefficients to be $10^{-5}$. We terminate our simulation at the time of $t_{\rm r} = 7.2t_{\rm A}$, the snapshot of which is adopted to explore test particle acceleration.

\subsection{Test particle simulations}\label{TestMethod}

The equation of motion for a charged particle is given by
\begin{equation}
 \frac{d}{d t} \left( \gamma m \vc{u} \right) = q \left( \vc{E} + \vc{u} \times
\vc{B} \right) , \label{eq:ptraj1}
\end{equation}
where $\gamma \equiv \left( 1 - u^2 / c^2 \right)^{-1/2}$ is the Lorentz factor of relativistic particle, and $c$ is the speed of light. The symbols $\vc{u}$, $m$, and $q$ are the particle velocity, mass, and electric charge, respectively. We focus on the acceleration processes resulting from the motional electric field $-\vc{v}\times \vc{B}$, ignoring the possible acceleration from resistivity effects of the electric field, i.e., the last term in Equation (\ref{eq:elec}). It has been demonstrated that the Fermi-type acceleration by the motional electric field dominates over the acceleration by non-ideal electric fields \citep{Guo:2019}. In our study, we make sure that the resistivity effects of the electric field do not enter, as $\eta$ could also be from numerical resistivity. Therefore, Equation (\ref{eq:ptraj1}) is further rewritten as 
\begin{equation}
 \frac{d}{d t} \left( \gamma m \vc{u} \right) = q \left[ \left( \vc{u} - \vc{v}
\right) \times \vc{B} \right] . \label{eq:ptraj2}
\end{equation}
With this equation, we integrate particle trajectories using the 8$^{th}$ order embedded Dormand-Prince method \citep{Hairer_etal:2008} with an adaptive time step based on an error estimator.\footnote{In terms of the precision of long-term integration, we explore different methods for integrating particle trajectories such as a simple functional iteration, the implicit Gauss-Legendre method, and the explicit Runge-Kutta method. After taking into the numerical accuracy and calculation time account, we choose the 8th-order explicit Runge-Kutta method by Dorman \& Prince with step control and dense output, using a very small precise control parameter of $10^{-8}$ to ensure high-precision numerical output. This method has a higher accuracy than the classic 4th-order explicit Runge-Kutta method and has also been used in recent publications \citep{Kowal:2021,ZhangXiang:2021}}
The local values of the plasma velocity $\vc{v}$ and magnetic field $\vc{B}$ at each step of the integration were obtained by cubic interpolation with the discontinuity detector based on a total variation diminishing limiter. 

In our simulations, for our results not to be limited to a particular astrophysical system, all physical quantities are normalized to be dimensionless, and the results are applicable to studying the acceleration of relativistic particles in the nonrelativistic reconnection process. We consider the relativistic test particles with their speed close to the light speed $c$ that is larger than the Alfv\'en speed $V_{\rm A}$ by several orders of magnitude. Therefore, we have $c\gg V_{\rm A} > v_{\rm rec}$, where $v_{\rm rec}$ is the velocity of turbulent magnetic reconnection. Since for a nonrelativistic reconnection process, the reconnection timescale is longer than the Alfv\'en wave crossing time ($t_{\rm A}$), which is much longer than the crossing time of a relativistic particle, we adopt a single snapshot from our reconnection simulation to perform the test particle simulations. In practice, with 10,000 protons as test particles that are injected instantly into a snapshot at $t_{\rm r}=7.2t_{\rm A}$, we integrate the evolution equation of particles over 0.1 in units of $t_{\rm A}$.\footnote{As in all numerical experiments, an optimal number of particles should be used. Our results presented below will not be changed with the number of particles if this number exceeds 5,000. Moreover, we find that after $t=6.0 t_{\rm A}$, the particle acceleration from one snapshot provides a representative result for all snapshots.}

\begin{figure*}
\centering
\includegraphics[width=0.45\textwidth,height=0.25\textheight]{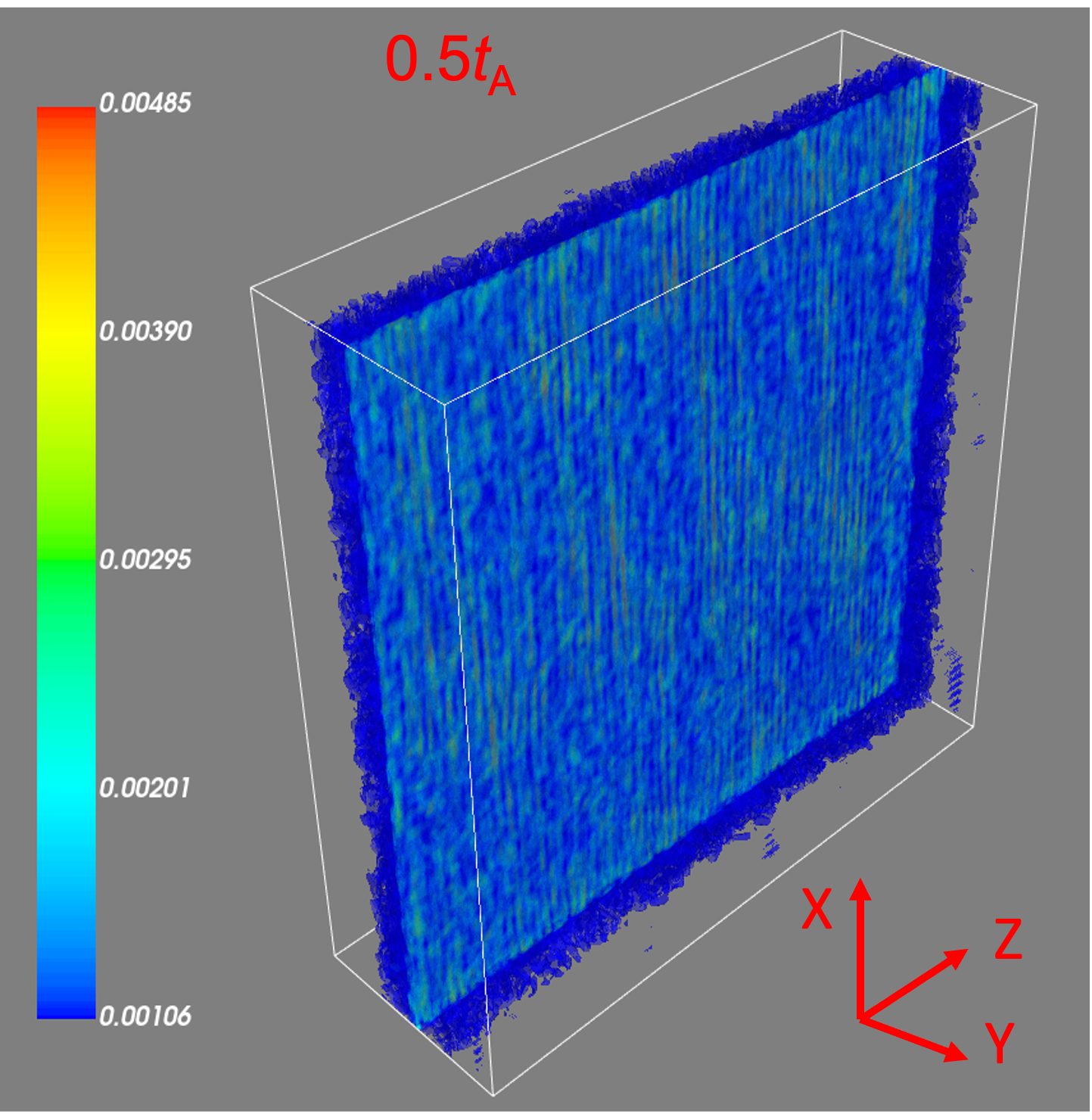} 
\includegraphics[width=0.45\textwidth,height=0.25\textheight,bb=0 40 450 330]{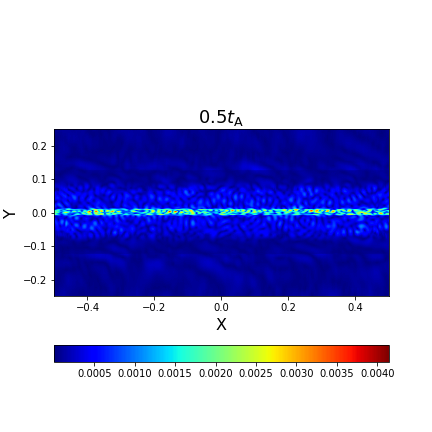}\ \
\includegraphics[width=0.45\textwidth,height=0.25\textheight]{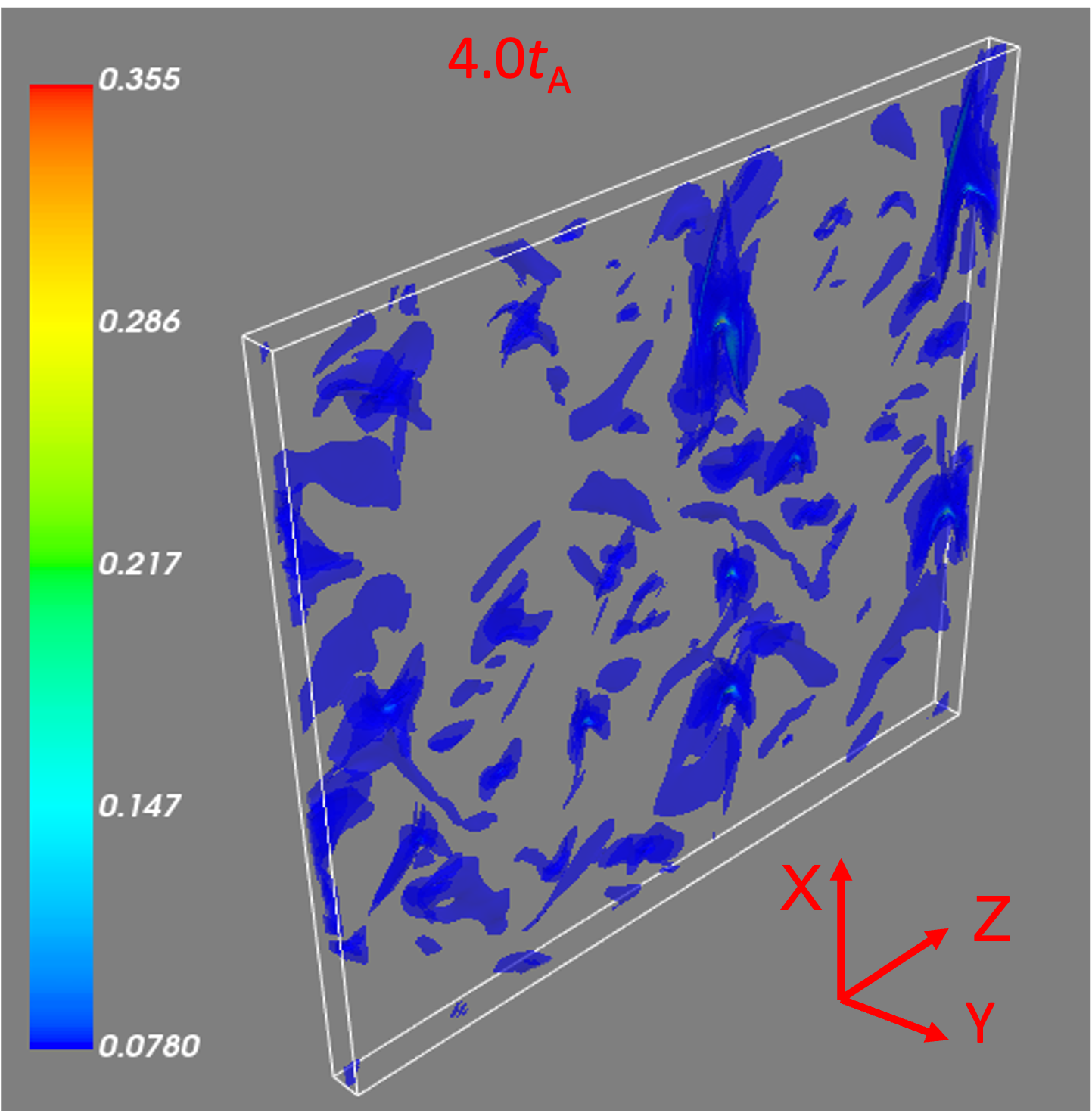}
\includegraphics[width=0.45\textwidth,height=0.25\textheight,bb=0 40 450 330]{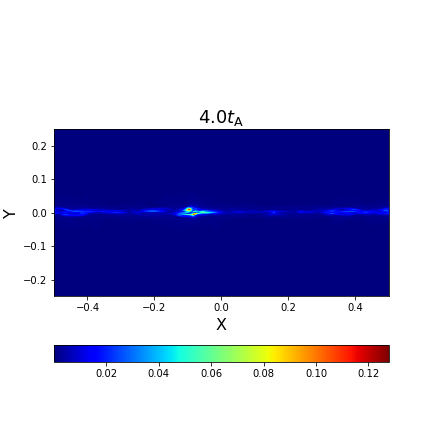} \ \
\includegraphics[width=0.45\textwidth,height=0.25\textheight]{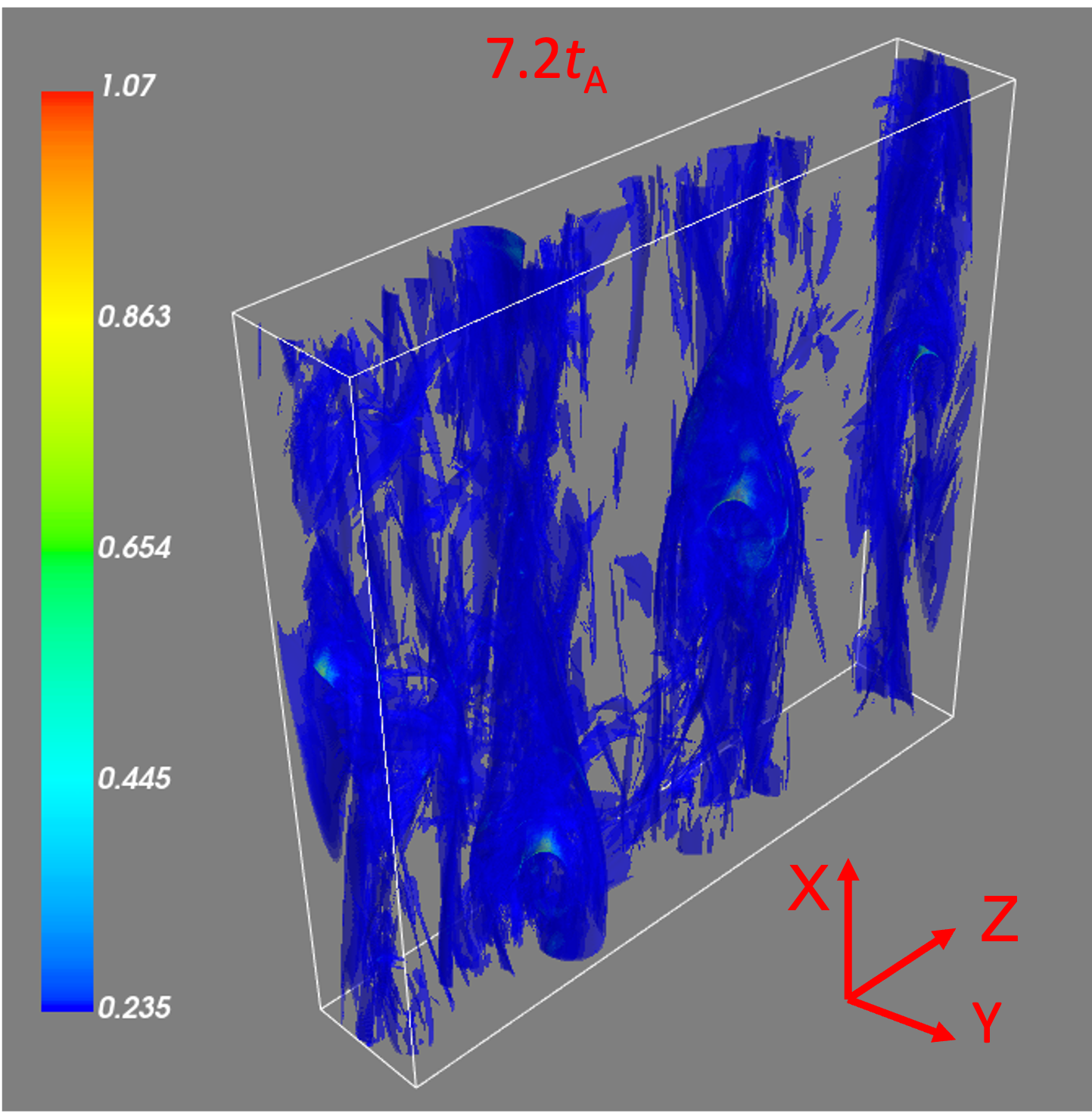} 
\includegraphics[width=0.45\textwidth,height=0.25\textheight,bb=0 40 450 330]{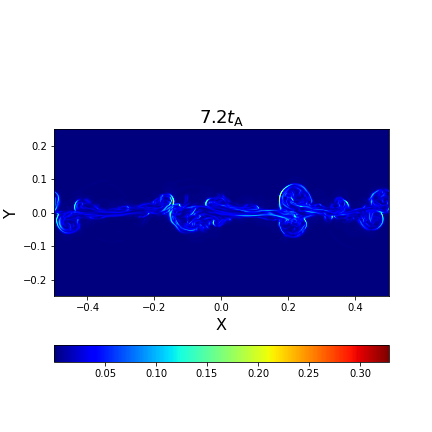}\ \
\caption{3D distributions of current densities (left column) and corresponding 2D imaging on the $X$-$Y$ plane (right column) at different times $t_{\rm r}\simeq$0.5, 4.0 and 7.2$t_{\rm A}$ (from upper to lower rows) during the evolution of reconnection.
} \label{fig:currdis2D}
\end{figure*}

\begin{figure*}
\centering
\includegraphics[width=0.99\textwidth,height=0.40\textheight]{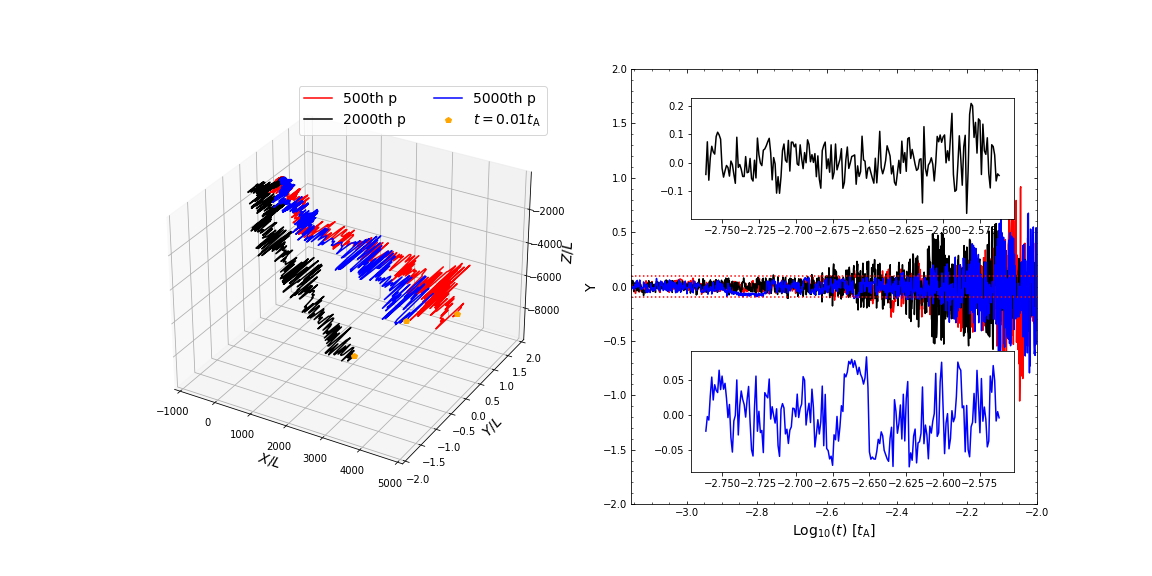} %,,bb=200 10 600 700
\caption{Left panel: the 3D trajectories of three test particles chosen from the test particle simulation. The plotted pentagrams indicate their locations at $t=0.1t_{\rm A}$. Right panel: the motions of particles in the direction of inflow directions (along the $Y$ axis) as a function of time, respectively corresponding to the particle trajectories shown in the left panel (one-to-one correspondence by the color of the curve). The horizontal dotted lines indicate the estimated boundaries of the reconnection layer. 
} \label{fig:3Dtraj}
\end{figure*}

\begin{figure*}
\centering
\includegraphics[width=0.99\textwidth,bb=80 80 1080 1020]{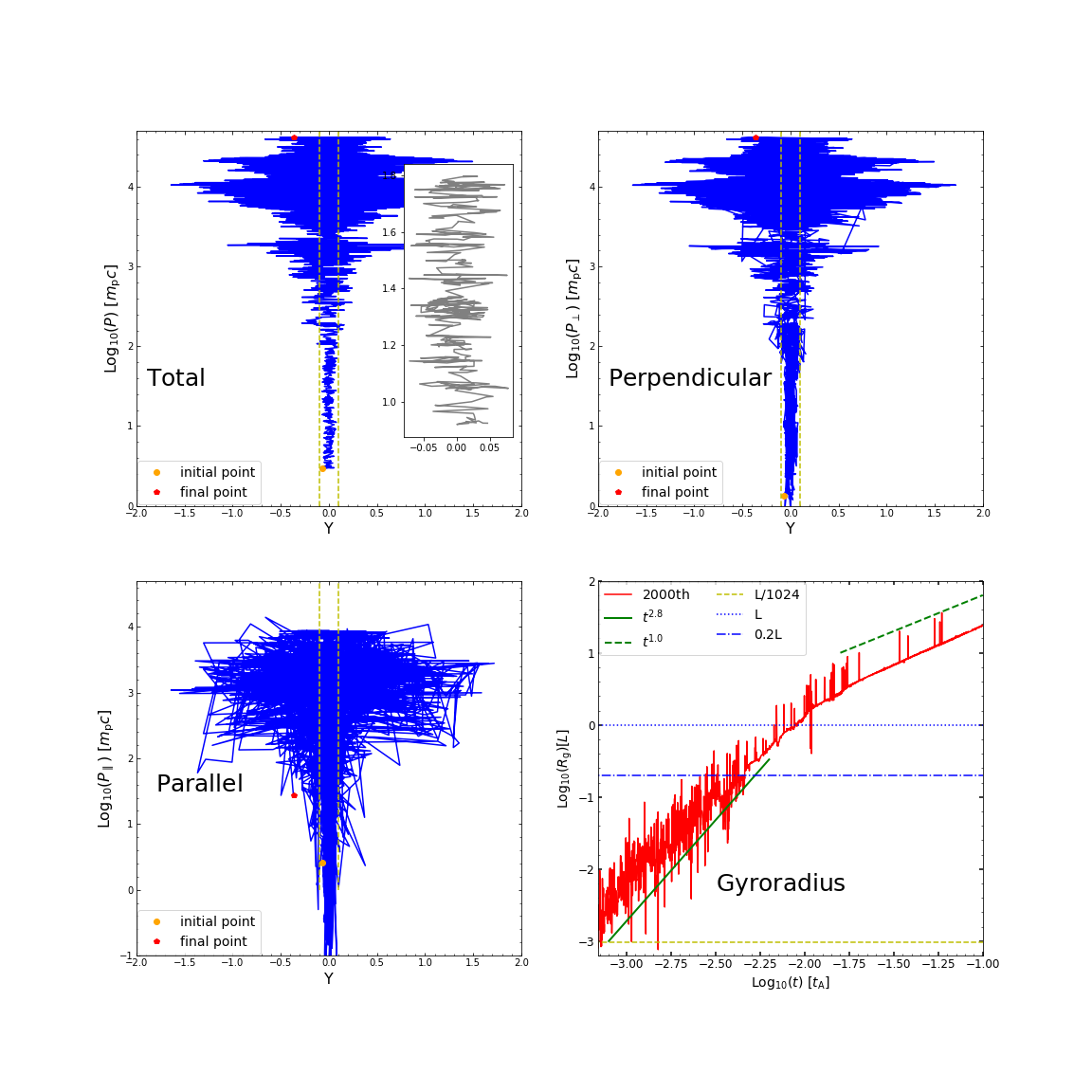}%,bb=0 120 1300 1200
\caption{Total momentum (left upper panel), its perpendicular (right upper panel), and parallel (left lower panel) components of a test particle vs. the location of the particle in the $Y$ direction. The inset in the left upper panel shows zoom-in. The initial and final positions of the particle are marked in each panel. The vertical dashed lines indicate the estimated range of the reconnection layer, $-0.1 \lesssim Y \lesssim 0.1$. The right lower panel indicates the gyroradius $R_{\rm g}$ vs. $t$ of the test particle. The horizontal dashed, dash-dotted, and dotted lines correspond to the grid size, the estimated thickness of the reconnection layer, and the box size, respectively. The green solid and dash-dotted lines show $R_{\rm g} \propto t^{2.8}$ and $R_{\rm g} \propto t^{1.0}$ as a reference, respectively.
} \label{fig:traj-turmr}
\end{figure*}

\section{Reconnection acceleration with reconnection-driven turbulence
}\label{Results1}

\subsection{Structures of turbulent reconnection layer}

Before studying the acceleration of the test particles, we in this subsection first explore the evolution of the current sheet structure. With a high-resolution simulation of $1024\times4096\times1024$, we explore how the current sheet evolves during the spontaneously driven turbulent reconnection process. At the selected three snapshots $t_{\rm r}\simeq$0.5, 4.0 and 7.2\ $t_{\rm A}$, we present 3D distributions of the current density (see left column of Figure \ref{fig:currdis2D}) and 2D slices of current density on the $X$-$Y$ plane at $Z=0$ (see the right column of Figure \ref{fig:currdis2D}). Here, the current density is obtained by $\vc{J}=\nabla \times \vc{B}$. As seen, the current density with a noise-like structure is evenly distributed in a sheet-like plate at the early stage of evolution ($\simeq0.5 t_{\rm A}$). It then begins to form high current density clumps at $t_{\rm r}\simeq 2.5 t_{\rm A}$ (not shown in the paper). When the system evolves to $t_{\rm r}\simeq4.0 t_{\rm A}$, we see a very inhomogeneous distribution of the current density. This is probably because the initial stochastic noise triggers various MHD instabilities (e.g., Kelvin-Helmholtz and tearing instabilities) that make the current sheet fragment. As the system evolves further, the current sheet begins to thicken and forms a significant magnetic flux rope with a multi-scale complex structure at $t_{\rm r}\simeq5.0 t_{\rm A}$ (not shown). With the periodic conditions set along the outflow directions, outflows do not leave the system, which leads to a continuous accumulation of energy. At later times, the reconnection layer is slowly eating through the undisturbed fluid, and reconnection-driven turbulence is fueled by the free energy of the oppositely directed magnetic fields. This will cause the current sheet to slowly widen and increase the volume of the reconnection layer with a lower growth rate. The detailed statistical analysis of the reconnection simulation demonstrates that MHD turbulence with Kolmogorov turbulent energy spectrum and scale-dependent turbulence anisotropy (GS95, LV99) is well developed at $t\gtrsim 5 t_{\rm A}$. Therefore, we stop our simulation at $t_{\rm r}=7.2 t_{\rm A}$. 

With a similar setup, \cite{Kowal_etal:2017} discussed the boundary effect on reconnection (see also \cite{Beresnyak:2017} for periodic conditions in the three directions). Recently, our reconnection-driven turbulence simulation by MHD-PIC considered a reflective boundary in the direction perpendicular to the current sheet and found that a reflective boundary makes larger-scale reconnection structures \citep{Liang2023}. Nevertheless, the problem of periodic boundary conditions will be further investigated and quantified in future work. As shown in Figure \ref{fig:currdis2D}, we observe that the evolution of stochastic reconnection leads to the formation of thickening regions in the current density. This broadened reconnection layer over the range $-0.1 \lesssim Y \lesssim 0.1$ provides a favorable place for the rapid acceleration of particles. The acceleration efficiency of particles in reconnection strongly depends on the reconnection rate, i.e., the inflow speed driven by reconnection \citep{Xu:2023}, which can be measured as the growth speed of the current layer width, i.e., $v_{\rm rec}=d\bigtriangleup/dt\simeq0.02V_{\rm A}$ at $t_{\rm r}=7.2 t_{\rm A}$ \citep[see][for more details]{Beresnyak:2017,Kowal_etal:2017}. Although the width of the reconnection layer increases constantly, the reconnection rate almost remains its stable value, which is similar to the results in \cite{Kowal_etal:2017} and \cite{Beresnyak:2017}.

%In general, the evolution process of the reconnection layer can be divided into three main phases: random noise fluctuation and amplification in a uniform current sheet ($t_{\rm r}\lesssim0.5 t_{\rm A}$); a nonuniform current sheet ($t_{\rm r}\simeq 2.5 t_{\rm A}$) and its fragmentation ($t_{\rm r}\simeq 4.0 t_{\rm A}$); and development of multi-scale turbulence and thickening of the reconnection layer ($t_{\rm r}\gtrsim 5.0 t_{\rm A}$), with a steady state reached at $t_{\rm r} \approx 6 t_{\rm A}$.  

\subsection{Reconnection acceleration}\label{hr}

\subsubsection{Trajectory analysis of particles}
With 3D data cubes taken from the above-mentioned reconnection simulation at the final snapshot of $7.2t_{\rm A}$, we perform test particle simulations to study the acceleration of particles that interact with reconnecting turbulent magnetic fields. As an example, the left panel of Figure \ref{fig:3Dtraj} presents the trajectories of three test particles with the almost same initial location inside the reconnection layer. As shown in this panel, the particles are confined in the vicinity of the reconnection region ($-0.1 \lesssim Y \lesssim 0.1$) to gain their kinetic energy by repeatedly bouncing back and forth off the inflows. 
 
The right panel of Figure \ref{fig:3Dtraj} further illustrates the motions of particles in the direction of the $Y$ axis (along the inflow directions) as a function of time, in which we also show two zoomed-in parts to illustrate the more details. The boundaries of the reconnection layer are indicated by the horizontal dotted lines. It can be seen that with the increase of particle energy, the acceleration proceeds with the particle moving beyond the reconnection layer after $t\approx 0.005t_{\rm A}$.

\subsubsection{Particle acceleration processes}
When particles are confined within and near the reconnection layer, they are accelerated with continuous energy gain. As an example, Figure \ref{fig:traj-turmr} shows the total $P$ (left upper), perpendicular $P_{\perp}$ (right upper), and parallel $P_\|$ (left lower) momentum of a test particle (normalized by $m_{\rm p}c$, where $m_{\rm p}$ is the proton mass) as a function of its location in the $Y$ direction. The perpendicular and parallel components of momentum are measured with respect to the local magnetic field.
The acceleration during turbulent reconnection leads to an energy increase of about 3 orders of magnitude in a zig-zag manner. The inset is a zoom-in to clearly show that the particle gains energy via bouncing back and forth crossing the reconnection layer many times. As argued in \cite{Xu:2023}, the kinetic energy of the reconnection-driven inflows is transferred to particles via their repeated head-on collisions. The lower right panel shows the distributions of the gyroradius of the particle, $R_{\rm g} \propto P_\perp/|q|B \propto t^{2.8}$, where $P_\perp$ is the particle perpendicular momentum. With the increase of particle energy, $R_{\rm g}$ generally increases with time. 
Its large fluctuations are mainly caused by both the large fluctuations of the magnetic fields (mainly $B_{\rm y}$ and $B_{\rm z}$ components) and velocities in the reconnection region. 

\begin{figure*}
\centering
\includegraphics[width=0.99\textwidth,bb=80 80 1080 900]{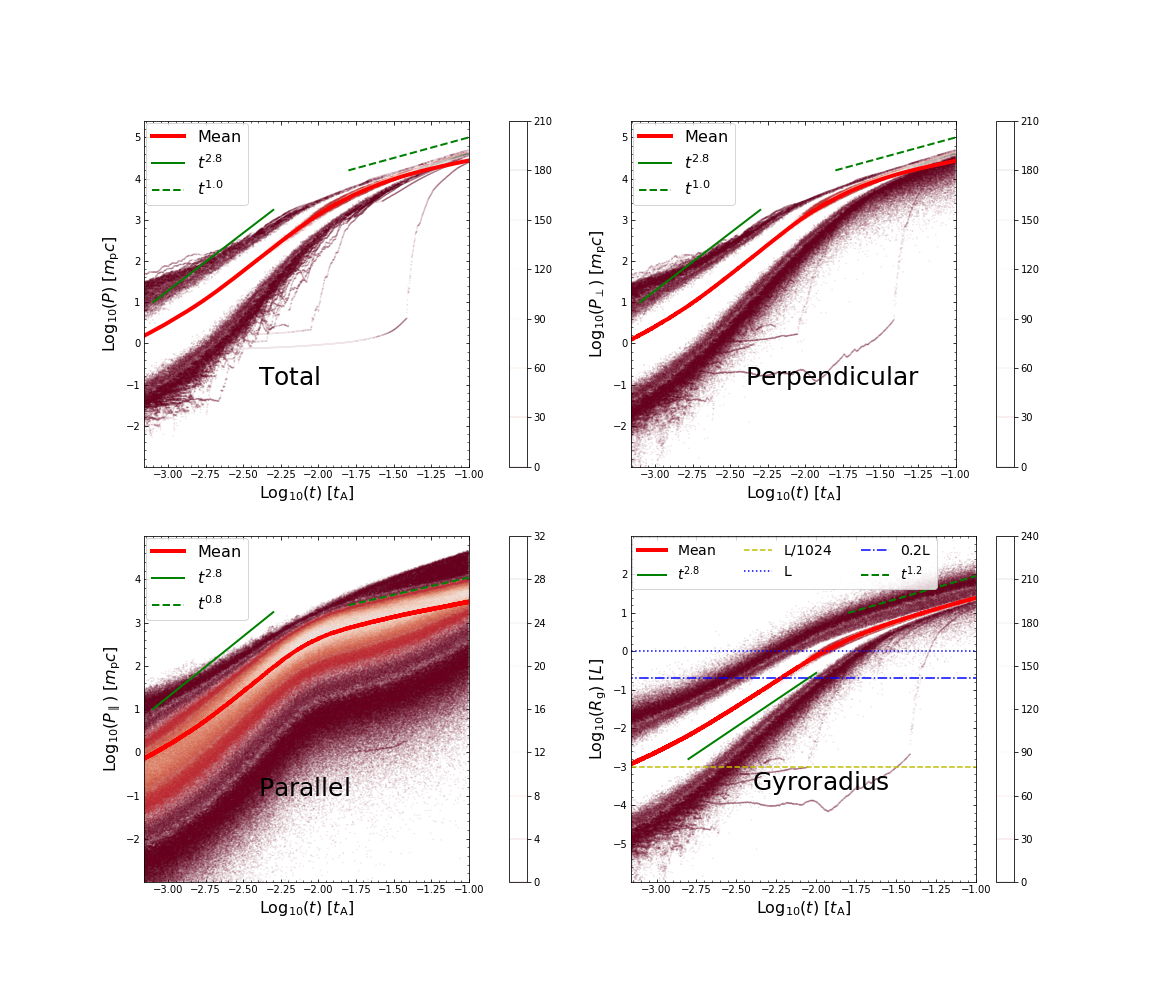}%,bb=20 20 400 350,bb=80 80 1080 550
\caption{Same as Figure \ref{fig:traj-turmr} but for 10,000 particles. The color scale indicates the particle distributions at different integration times in units of $t_{\rm A}$, and the red lines represent the mean values. 
} \label{fig:tdis-turmr}
\end{figure*}

The total $P$, perpendicular $P_\perp$ and parallel $P_\parallel$ momentum, and gyroradius $R_{\rm g}$ of all 10,000 test particles as a function of the time $t$ are shown in Figure \ref{fig:tdis-turmr}, where the red thick solid lines represent the mean values and the color scales indicate their distribution. In the right lower panel, the horizontal dotted and dashed lines indicate the box scale $L$ in the $X$ and $Z$ directions, and the grid size of $h=L/1024$, respectively, while the blue dash-dotted line indicates the approximate thickness of the reconnection layer. As shown, the momentum and gyroradius of particles all increase with time, roughly following $\propto t^{2.8}$, with the energy increase by about 3 orders of magnitude in the range of box size. The increase of both parallel and perpendicular momentum of particles is consistent with the theoretical expectation in \cite{Xu:2023} and previous simulations \cite[e.g.,][]{Kowal_etal:2011,delValle_etal:2016}. When the gyroradius of particles is greater than the box size, we see a shallower power law of approximately $P\propto t^{1.0}$ ($R_{\rm g} \propto t^{1.2}$) due to our periodic boundary conditions. During this stage, the gain of energy is due to a much slower drift acceleration caused by gradients of large-scale magnetic fields in the perpendicular direction 
(see the right upper panel). The presence of a guide field allows the particles to slightly accelerate in the parallel direction as well (see the left lower panel for $P_\parallel \simeq 0.1 P_\perp$).

As shown in Figures \ref{fig:traj-turmr} and \ref{fig:tdis-turmr}, during the exponential growth phase with the power law of $\propto t^{2.8}$, the gyroradius of the accelerated particles becomes comparable to the size of the box. To confine the particles around the reconnection layer, the Larmor radius of particles should be smaller than the box size $L$, where the maximum energy can be estimated by $E_{\rm max} \propto qLB$. However, due to the setting of periodic boundary conditions, the particles can be continuously accelerated with a lower acceleration efficiency (with $R_{\rm g} \propto t^{1.2}$) after $R_{\rm g}\gtrsim L$. We would like to stress that the energy increase by three orders of magnitude seen in the simulation is limited by the box size, corresponding to three orders of magnitude of the particle gyroradius, beyond which the energy increase is due to the consideration of periodic boundary conditions.

 Our simulations show that particles with both $R_{\rm g}$ smaller and larger than the thickness of the reconnection layer ($\sim 0.2 L$) can be accelerated. In the former case, particles bounce back and forth within the reconnection layer, while in the latter case, particles gyrate around the reconnection layer. Theoretical models for turbulent reconnection acceleration in these two different cases can be found in \cite{deGouveiaLazarian:2005,Lazarian:2012ASSP,Xu:2023} (see also Figure \ref{fig:3Dtraj}). The numerical demonstration has been carried out by MHD simulations for the former case \cite{Kowal_etal:2011,Kowal_etal:2012b,delValle_etal:2016} and by kinetic simulations for the latter case \cite{ZhangSironi:2021,ZhangQL:2021}. Here, we clearly see both cases in a self-driven reconnection for the first time (to our knowledge).

\begin{figure*}
\centering
\includegraphics[width=0.99\textwidth,height=0.35\textheight,bb=90 50 980 490]{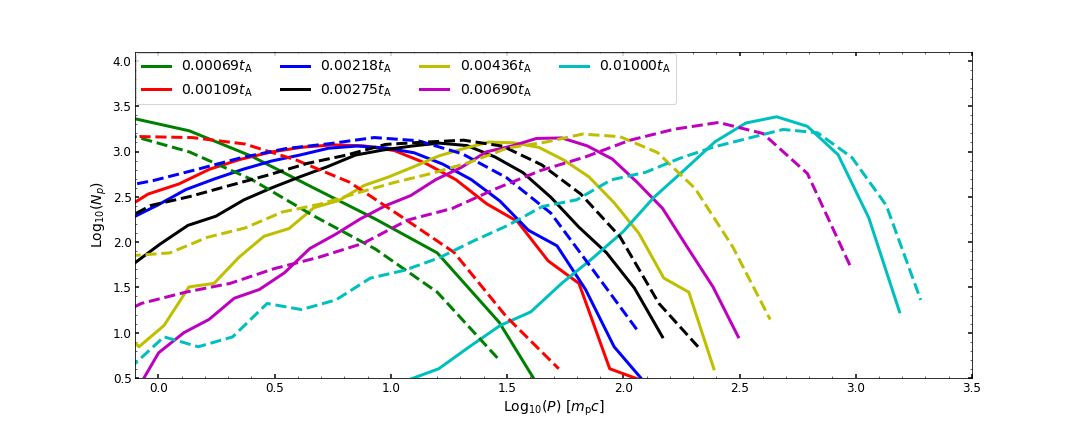}\ \ %,bb=20 20 400 350
\includegraphics[width=0.99\textwidth,height=0.35\textheight,bb=90 10 980 440]{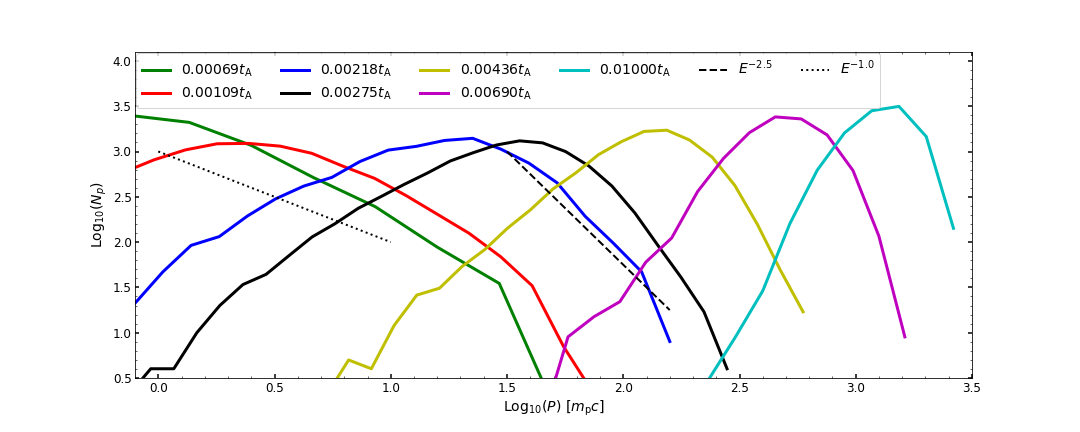}
\caption{Spectral energy distributions of particles accelerated in the self-driven turbulent reconnection at different particle integration times in units of $t_{\rm A}$, for the perpendicular (solid lines) and parallel (dashed lines) components of the momentum (upper panel), and for the total momentum (lower panel). In the upper panel, the same color represents the perpendicular and parallel distributions at the same integration time.} 
\label{fig:sed-turmr}
\end{figure*}

\subsubsection{Spectral energy distribution of accelerated particles}
Spectral energy distributions of accelerated particles are plotted in Figure \ref{fig:sed-turmr} at the integration time $t\simeq 0.00069 t_{\rm A}$, $0.00109 t_{\rm A}$, $0.00218 t_{\rm A}$, $00275 t_{\rm A}$, $0.00436 t_{\rm A}$, $0.00690 t_{\rm A}$, and $0.01000 t_{\rm A}$, which correspond to the acceleration processes within and around the reconnection layer. The upper panel of this figure shows the spectra of momentum components perpendicular and parallel to the local magnetic field by the solid and dashed lines, respectively. Note that the particles are injected with an initial Gaussian distribution of energies, which are decomposed into perpendicular and parallel components. With increasing integration time, as a result of reconnection acceleration, particle energy spectra shift to higher and higher energies. Spectral continuously broadening is not very clear as we use instant injection. As seen, the acceleration in the perpendicular direction dominates an increase of particle energy only at an early time before about $0.002t_{\rm A}$. Most time, the acceleration in the parallel direction dominates the energy increase. 

The lower panel of Figure \ref{fig:sed-turmr} presents the time evolution of the spectrum of total particle momentum. Most time the total energy spectrum follows that of the perpendicular energy. The energy distribution of accelerated particles exhibits a non-thermal tail which is estimated as a power-law distribution of $N_{\rm p}(P)\propto P^{-\alpha}$, where the spectral index $\alpha$ evolves from approximately 1.0 to 2.5. The slope 2.5 is consistent with the theoretical expectation in \cite{Xu:2023} for nonrelativistic reconnection acceleration with a weak guide field \cite[see also][]{deGouveiaLazarian:2005}. Here, we do not consider the escape of particles from the box in our simulations. This setup is applicable to the case when the escape from the entire system happens much slower than the acceleration. Indeed, we observe that during our simulation, particles are mainly confined in and around the reconnection layer (see Figure \ref{fig:3Dtraj}). So the escape from the entire system is relatively slow. As we do not continuously inject new particles at low energies, with the acceleration of the bulk population, the minimum energy and the peak of the energy distribution both move to higher and higher energies with the increase of simulation time. During the exponential growth process of $\propto t^{2.8}$ that we are interested in, the energy particles can reach is limited by our box size. As a result, it is difficult to see an extended power-law tail of the accelerated particles. During the early time of the particle simulation, a power-law shape can be seen over approximately one order of magnitude in energies. 

As here we deal with the acceleration process with spatial inhomogeneity (see Figure \ref{fig:currdis2D}). The ``escape” can happen locally from a region with a higher level of turbulence and higher reconnection efficiency (and thus higher acceleration efficiency) to a region with a lower level of turbulence and lower reconnection efficiency (and thus lower acceleration efficiency). This local ``escape” may affect the measured spectral shape. We note that the local ``escape” can be time-dependent with the increase of particle gyroradius from a value much smaller than the largest thickness of the reconnection layer to a value larger than it.

The particle energy spectra we obtained are similar to those obtained in other particle-in-cell (PIC) and MHD simulations  \cite[e.g.,][]{Cerutti_etal:2013,Guo_etal:2014,Kowal_etal:2012,Medina:2021,Pezzi:2022}. The non-thermal power-law tail cannot further extend to higher energies. In realistic astrophysical systems with a limited size, e.g., in the black hole X-ray binary jets, the direct solution of the Fokker Planck equation controlling electron evolution also predicts a non-thermal tail with non-extending power-law features \cite{Zhang:2018}. In the former case considered here, the smallest energy of particles increases with time. As a result, we see that in the late stage of particle acceleration, the spectral distribution of particles becomes narrower as shown in Figure \ref{fig:sed-turmr}.

\section{Discussion}\label{Diss}

In the framework of the {\it externally driven turbulent reconnection}, \cite{Kowal_etal:2012} demonstrated that charged particles trapped within the reconnection layer experience many head-on collisions 
with contracting magnetic fields which significantly enhances the acceleration rate. Differently, our present numerical work is an attempt to explore the properties of particle acceleration with self-driven turbulent reconnection. It is noted that in the framework of the externally driven turbulent reconnection, particle energy spectral distributions presented in \cite{delValle_etal:2016} are similar to our current analysis. Although the understanding of CR acceleration in various MHD turbulence settings has been discussed by different authors \citep{Lehe:2009,Lynn:2013,Guo_etal:2014,Guo:2019,Ergun:2020,Guo:2021,ZhangQL:2021,ZhangXiang:2021,Isliker:2022}, the acceleration by self-driven turbulent reconnection is an insufficiently studied subject. 

Tearing reconnection has been considered an alternative mechanism for the fast magnetic reconnection process. The particle acceleration in magnetic islands (ropes) has been extensively studied using 2D PIC simulations of collisionless electron-ion or electron-positron plasmas  \cite[e.g.,][]{ZenitaniHoshino:2001,Drake_etal:2006,Drake_etal:2010,LyubarskyLiverts:2008,ClausenLyutikov:2012,Cerutti_etal:2013,Li_etal:2015}, and also using  3D PIC simulations \cite[e.g.,][]{SironiSpitkovsky:2014,Cerutti_etal:2014,Guo_etal:2014,ZhangQL:2021,Nathanail:2022}. It should be pointed out that these studies can only probe particle acceleration at the kinetic scales of the plasma, i.e., a few hundredths of the skin depth. The current difficulty in achieving sufficiently large length scales is the reason why the 3D PIC simulations do not observe turbulent flow behavior in the reconnection layer \citep{Lazarian_etal:2020}. On the other hand, some turbulence-related processes, such as flux-freezing violation \citep{Eyink:2011} and Richardson dispersion \citep{Richardson:1926,Eyink_etal:2013} that follow turbulent reconnection theory (LV99), can not be explained in the framework of the tearing reconnection. The numerical results of the turbulent reconnection provided in \cite{Kowal_etal:2020} demonstrate that the tearing mode plays a role only at the early stage of reconnection. As the reconnection layer evolves in time, the tearing instability is suppressed and the Kelvin–Helmholtz instability plays the dominant role in driving turbulence and initiating the turbulent reconnection. 

Compared with \cite{Kowal_etal:2017} in the case of externally driven turbulent reconnection, we find that in the present reconnection-driven turbulence, the phase of exponential acceleration with an index of $\sim 2.8$ shown in Figure \ref{fig:tdis-turmr} is similar to the earlier low-resolution simulations with index approximately $2.7$. This exponential acceleration process is related to the largest thickness of the reconnection layer of $\sim 0.2L$. The turbulence driven by the reconnection in this paper has very different injection properties compared to that of \cite{Kowal_etal:2017}. The former happens at a large scale by an external force, and the latter due to the action of the reconnection itself has a different injection caused by the associated instabilities from the initial small-scale perturbations. 

As turbulence regulates the reconnection rate, the reconnection rate with externally-driven turbulence and reconnection-driven turbulence can be very different \cite{Kowal_etal:2017}. The efficiency of reconnection acceleration associated with the motional electric field induced by inflows depends on the reconnection rate \cite{Xu:2023}. Therefore, we expect that the acceleration efficiency has dependent on the turbulence-driving mechanism in reconnection. Despite those differences, we find that the overall acceleration process is very similar in terms of energy growth. As a result, reconnection-driven turbulence can effectively accelerate particles.

We have explored whether particle acceleration takes place rather than studied the evolution of the spectrum with the evolution of the turbulent reconnection layer. For this purpose, the snapshot of the simulations is reasonable. Note that the frozen-box approximation in 2D MHD simulation has been claimed to lead to a super-Fermi acceleration \cite{Majeski:2023}. In the case of a 3D simulation, whether there is an artificial super-Fermi acceleration requires specialized research in the future. As for reconnection acceleration, it would be more desirable to self-consistently explore the acceleration of particles together with the co-evolution of fluids, as done by e.g., \cite{Lehe:2009} and \cite{Lynn:2013} for particle acceleration in MHD turbulence. In general, the PIC and MHD simulations are complementary for studying particle accelerations. The obvious advantage of MHD simulations is that they can simulate turbulent reconnection on macroscopic scales which is challenging with the PIC.

\section{Summary}\label{Summ}
We have numerically studied for the first time particle acceleration in the magnetic reconnection with reconnection-driven turbulence.\footnote{Around the same time, \cite{Liang2023} used an MHD-PIC method to simulate reconnection-driven turbulence and turbulent reconnection acceleration of particles. They found that particles can be efficiently accelerated in reconnection-driven turbulence.} Our research confirmed the theoretical picture proposed by \cite{deGouveiaLazarian:2005}
and further developed and quantified in \cite{Xu:2023} that the particles can be efficiently
accelerated within the turbulent reconnection layer via bouncing back and forth between converging magnetic fields. The current numerical study with high-resolution simulations on reconnection acceleration with self-driven turbulence extends earlier numerical studies with low resolution and externally driven turbulence \cite{Kowal_etal:2011,Kowal_etal:2012,delValle_etal:2016}. 

The main results are summarized as follows:

\begin{itemize}
\item The self-driven turbulence is important not only for enabling the fast turbulent reconnection but also for enabling the particle acceleration in the turbulence-broadened reconnection layer, with the energy gain coming from the kinetic energy of the reconnection-driven inflows. Therefore, the acceleration efficiency is expected to strongly depend on the reconnection rate. 

\item We found that the acceleration in the direction perpendicular to the local magnetic fields dominates that in the parallel direction, which is consistent with the theoretical expectation in \cite{Xu:2023}. The particle energy increases with time by about 3 orders of magnitude in the range of the box size and approximately follows the scaling of $P\propto t^{2.8}$ before $t\simeq 0.01 t_{\rm A}$. Given periodic boundary conditions, the particle energy can be increased constantly with the increase of simulation time and approximately follows the scaling of $P\propto t^{1.0}$.

\item We demonstrate the reconnection acceleration of particles with their gyroradii both smaller and larger than the thickness of the reconnection layer. The latter case was first theoretically predicted by \cite{Lazarian:2012ASSP}.

\item The energy spectra of the accelerated particles present the non-thermal power-law tail, $N_{\rm p} (P)\propto P^{-\alpha}$, with $\alpha$ evolving from $\sim 1$ to $\sim 2.5$. The upper limit slope $\sim 2.5$ is consistent with the theoretical expectation in \cite{Xu:2023} for nonrelativistic reconnection acceleration with a weak guide field.

\end{itemize}

Our current study demonstrated efficient particle acceleration in self-driven turbulent reconnection. It has important implications on a wide range of astrophysical problems, including reconnection acceleration in e.g., accretion discs, jets, and the origin of high-energy cosmic rays.

\section{Acknowledgements}
We thank the anonymous referee for valuable comments that significantly improved the quality of the paper. J.F.Z. acknowledges the support from the National Natural Science Foundation of China (grants No. 11973035), the Hunan Province Innovation Platform and Talent Plan--HuXiang Youth Talent Project (No. 2020RC3045), and the Hunan Natural Science Foundation for Distinguished Young Scholars (No. 2023JJ10039). G.K. thanks for support from the São Paulo Research Foundation, FAPESP (grants 2013/10559-5 and 2019/03301-8). A.L. thanks the support of NSF AST 1816234 and NASA TCAN 144AAG1967 and NASA AAH7546 grants.

\bibliography{ms}% Produces the bibliography via BibTeX.

\end{document}